\documentclass[prb]{revtex4}
\usepackage[ascii]{inputenc}
\usepackage[T1]{fontenc}
\usepackage[english]{babel}
\usepackage{amsmath,amssymb,amsfonts,textcomp}
\usepackage{color}
\usepackage{array}
\usepackage{supertabular}
\usepackage{hhline}
\usepackage{hyperref}
\hypersetup{pdftex, colorlinks=true, linkcolor=blue, citecolor=blue, filecolor=blue, pagecolor=blue, urlcolor=blue, pdftitle=Introduction, pdfauthor=ChineduEkuma, pdfsubject=, pdfkeywords=}
\usepackage[pdftex]{graphicx}
\newcommand\textsubscript[1]{\ensuremath{{}_{\text{#1}}}}

\newcommand\textstyleStrongEmphasis[1]{\textbf{#1}}
\newcommand\textstyleEmphasis[1]{\textit{#1}}
\makeatletter
\newcommand\arraybslash{\let\\\@arraycr}
\makeatother
\newcommand\liststyleWWviiiNumii{%
\renewcommand\theenumi{\arabic{enumi}}
\renewcommand\theenumii{\alph{enumii}}
\renewcommand\theenumiii{\roman{enumiii}}
\renewcommand\theenumiv{\arabic{enumiv}}
\renewcommand\labelenumi{\theenumi.}
\renewcommand\labelenumii{\theenumii.}
\renewcommand\labelenumiii{\theenumiii.}
\renewcommand\labelenumiv{\theenumiv.}
}
\makeatletter
\newcommand\ps@Standard{
  \renewcommand\@oddhead{}
  \renewcommand\@evenhead{\@oddhead}
  \renewcommand\@oddfoot{\thepage{}}
  \renewcommand\@evenfoot{\@oddfoot}
  \renewcommand\thepage{\arabic{page}}
}
\newcommand\ps@FirstPage{
  \renewcommand\@oddhead{}
  \renewcommand\@evenhead{\@oddhead}
  \renewcommand\@oddfoot{\thepage{}}
  \renewcommand\@evenfoot{\@oddfoot}
  \renewcommand\thepage{\arabic{page}}
}
\makeatother
\title{Introduction}
\begin{document}
\clearpage\setcounter{page}{1}\pagestyle{Standard}
\thispagestyle{FirstPage}
{\centering
\textbf{First Principle Local Density Approximation Description of the
Electronic Properties of Ferroelectric Sodium Nitrite}
\par}

\bigskip

{\centering
C. E. Ekuma, M. Jarrell, and J. Moreno \\
 Departmentof Physics \& Astronomy, and Center for Computation \& Technology, Louisiana State University (LSU)\\
 Baton Rouge, Louisiana 70803, USA
\par}

\bigskip

{\centering
L. Franklin, G. L. Zhao, J. T. Wang, and D. Bagayoko
\par}

{\centering
Department of Physics, Southern University and A\&M College in Baton
Rouge (SUBR) Baton Rouge, Louisiana 70813, USA
\par}

\bigskip

\bigskip

\noindent  The electronic structure of the ferroelectric crystal,
NaNO\textsubscript{2}, is studied by means of first-principles, local
density calculations. Our ab-initio, non-relativistic calculations
employed a local density functional approximation (LDA) potential and
the linear combination of atomic orbitals (LCAO). Following the
Bagayoko, Zhao, Williams, method, as enhanced by Ekuma and Franklin
(BZW-EF), we solved self-consistently both the Kohn-Sham equation and
the equation giving the ground state charge density in terms of the
wave functions of the occupied states. We found an indirect band gap
of 2.83 eV, from W to R. Our calculated direct gaps are 2.90,
2.98, 3.02, 3.22, and 3.51 eV at R, W, X, ${\Gamma}$, and
T, respectively. The band structure and density of states show
high localization, typical of a molecular solid. The partial density of
states shows that the valence bands are formed only by complex anionic
states. These results are in excellent agreement with experiment. So
are the calculated densities of states. Our calculated electron
effective masses of 1.18, 0.63, and 0.73 m\textsubscript{o }in the
${\Gamma}$-X, ${\Gamma}$--R, and ${\Gamma}$-W directions,
respectively, show the highly anisotropic nature of this material.

\bigskip

\textbf{Pacs Numbers: }77.84.-s, 71.20.-b, 71.20.Mq, 71.20.Nr

\bigskip

\section{Introduction and Motivation}
Sodium nitrite is one of the ABO\textsubscript{2} group with several
interesting properties such as ferroelectricity, paraelectricity and
piezoelectricity. The ground state of NaNO\textsubscript{2} has one of
the simplest structures in the ABO\textsubscript{2} group. Since the
discovery of its ferroelectric property in 1958, by Sawada et al. [1],
there has been increased interest in diverse properties of
NaNO\textsubscript{2}, ranging from structural changes to dielectric,
electronic, thermal, electrical and elastic properties, as well as its
non-linear optical spectra [2]. Despite this noted interest in
ferroelectric NaNO\textsubscript{2}, there has been very little study
of the electronic band structure of the crystal [3].

One of the earliest electronic band gap measurements of ferroelectric
NaNO\textsubscript{2 }was by Asao et al. [4] who found band gaps of
2.3, 2.6, 3.1, and 3.7 eV for different current directions in their
resistivity work. Also, using the same method, Takagi and Gesi [5]
found a band gap of 2.4 eV. These authors did not specify whether the
gaps in question are direct or indirect. The infrared absorption
spectra study of Sidman [6] at a temperature of 77 \textit{K} gave a
band gap of 3.22 eV while that of Verkhovskaya and Sonin [7], at 293.15
K, led to a value of 3.14 eV. The semi-empirical LCAO work of El-Dib
and Hassan [8], for room temperature (293.15 K), found a band gap of
3.24 eV. The scanning electron microscopy, diffuse reflectance
spectroscopy, and electrical measurements study of Balabinskaya et al.
[9], at a temperature of 300 \textit{K }, reported an absorption edge of 3.0
eV.

One of the earliest theoretical studies of the band structure of
ferroelectric NaNO\textsubscript{2} was the X${\alpha}$ (${\alpha}$ =
0.75) exchange calculation of Kam et al. [10] who found a direct band
gap of 3.45 eV at ${\Gamma}$ and an indirect band gap of 2.0 eV. The
Full-potential linear muffin-tin orbital (FP-LMTO) method of Ravindran
et al. [2] found an indirect band gap of 2.2 eV. The generalized
gradient approximation (GGA) work of Wang et al. [11] utilized a full-potential 
linearized augmented plane wave (FL-APW) method to calculate
a band gap of less than 2.1 eV. The LDA approach of Zhuravlev and
Korabel{\textquoteright}nikov [12], using the nonlocal
Troullier--Martins (TM) pseudopotentials and the Slater exchange
potential with a correlation correction, led to a direct band gap of
2.50 eV at \textit{W}. Zhuravlev and Poplavnoi [13], also using
nonlocal Troullier--Martins (TM) pseudopotentials and the Slater
exchange (${\alpha}$ = 1) potential with a correlation correction,
found a minimum forbidden band gap of 3.07 eV. The LDA result of
Jiang et al. [14], using an orthogonalized linear combination of atomic
orbitals (OLCAO) method, found an indirect band gap of 2.95 eV from
\textit{S to R} \ symmetry points. The periodic Hatree-Fock (PHF)
method of McCarthy [15] overestimated the gap by a factor of 2-3.
Henkel et al. [16] employed their
\textcolor[rgb]{0.2,0.2,0.2}{plane-wave and Gaussian basis set and an
}X${\alpha}$ exchange and reported a direct band gap of 2.70 eV. These
results suffice to see that theoretical calculations have not resolved
the under- or overestimation of the band gap of ferroelectric
NaNO\textsubscript{2}.

In light of the preceding overview, the aim of this study is to attempt
to obtain the measured, fundamental band gap and other electronic
properties of ferro-NaNO\textsubscript{2} with ab-initio,
self-consistent LDA calculations. The confirmation of
our predictions of the band gaps and other properties for
BaTiO\textsubscript{3} [17], TiO\textsubscript{2} [18], CdS [19],
c-Si\textsubscript{3}N\textsubscript{4} [20], c-InN [21], and
SrTiO\textsubscript{3} [22] is a basis for the above presumption. 
Further, the mathematical rigor of the method [18-24] suffices to
expect it to lead to much better results.

\bigskip

\section{Method of Calculation}

Our calculations employed the Ceperley and Alder [25] local density
functional potential as parameterized by Vosko, Wilk, and Nusair [26].
We implemented the linear combination of atomic orbitals (LCAO), using
Gaussian functions for the radial parts. We utilized a program package
developed at the Ames Laboratory of the US Department of Energy (DOE),
in Ames, Iowa [27]. Our calculations are non-relativistic and were
performed using low temperature (120 K) lattice constants. The
distinctive feature of our approach resides in the implementation of
the Bagayoko, Zhao, and Williams (BZW) method, as enhanced by Ekuma and
Franklin (BZW-EF), consisting of concomitantly solving
self-consistently two coupled equations as explained below. One of
these equations is the Schr\"odinger type equation of Kohn and Sham
[28], referred to as the Kohn-Sham (KS) equation. The second equation,
which can be thought of as a constraint on the KS equation, is the one
giving the ground state charge density in terms of the wave functions
of the occupied states.

The essentials of the method follow. Beginning with a minimum or small
basis set capable of accounting for all the electrons in the system
under study, one performs ab-initio, self-consistent calculations. This
minimum or small basis set is constructed using orbitals resulting from
self-consistent calculations for the atomic or ionic species in the
solid. As noted below, these species are Na\textsuperscript{1+},
N\textsuperscript{1+}, and O\textsuperscript{2-} for
NaNO\textsubscript{2}. Subsequently, this basis set is augmented
with one orbital and self-consistent calculations are done. This new
orbital and others that may be added later are outputs of the
self-consistent studies of the ionic species. The occupied energies of the two
calculations for the solid are compared numerically and graphically.
These occupied energies from Calculations I and II are generally
different. A third calculation is carried out after adding another
{\textquotedblleft}ionic{\textquotedblright} orbital to the basis set.
We note that adding an orbital means augmenting the size of the basis
set by 2, 6, 10, or 14, depending on the $s$, $p$, $d$, or $f$ character of it,
respectively. Again, the occupied energies from Calculations II and III
are compared. This process continues until the occupied energies from a
calculation, i.e., N, are found to be identical to those from
Calculation (N+1) immediately following it, within our computational
uncertainties of 50 meV or less. At that point, the calculations are
completed and the results from Calculation N represent the physical
description of the system under study. The basis set for Calculation N
is referred to as the optimal basis set. Calculation (N+1) and
others, with larger basis sets that contain the optimal one, provided
linear dependency is avoided, reproduce the occupied energies from
Calculation N. However, by virtue of the Rayleigh theorem [19,23],
these calculations produce some unoccupied eigenvalues that are lower
than those from Calculation N, due to a non-trivial basis set and
variational effect [23].

All the self-consistent calculations known to us carry out the iterative
procedure that involves both the Kohn-Sham equation and the equation
giving the ground state charge density. However, the methodical
increase of the size of the basis set, as done in the BZW-EF method,
entails changes in the radial and angular characteristics of the basis
set as well as an increase in its size. It is in this sense that the
BZW-EF method solves self-consistently the two equations in question to
obtain, in a verified fashion, the minima of the occupied energies. The
difference between our method and single trial basis set calculations
may be best understood by noting that any serious radial, angular, or basis
size deficiencies that may exist in a single trial basis set cannot be
remedied by the iterative process noted above. In the case of the
BZW-EF method, such deficiencies are corrected as the methodical
augmentation of the basis set is carried out. The attainment of the
minima of the occupied states signifies that there is no deficiency
left to correct. When a larger basis set that contains the optimal one
is utilized to carry out a calculation, the charge density, the
potential, and the Hamiltonian resulting from self-consistency, as well
as the occupied eigenvalues, are the same as the corresponding ones obtained
with the optimal basis set. Hence, the Rayleigh theorem applies to the
outputs of this calculations whose occupied energies are the same as
those obtained with the optimal basis set. However, some unoccupied
energies from the larger basis set are generally lower than (or equal) to their
corresponding ones obtained with the optimal basis set. The lowering of
these unoccupied energies cannot be ascribed to physical interactions,
as the Hamiltonian did not change. It results from the non-trivial
and well-defined basis set and variational effect.

The enhancement of the original BZW method is in the methodical
increment of the basis set. This enhancement leads to adding the
polarization (\textit{p, d, or f}) orbitals, for a given principal
quantum number, before adding the spherically symmetric \textit{s} 
orbital (see Table 1). These additional unoccupied orbitals are needed
to accommodate the reorganization of the electron cloud, including
possible polarization, in the crystal environment. For valence
electrons in molecules to solids, polarization has primacy over
spherical symmetry.

While there are no specific rules governing the order of adding
orbitals, our experience has been that orbitals often need to be added
to the heaviest element before the others. Among these others,
preliminary studies of charge transfer serve as a significant guide. In
the case of NaNO\textsubscript{2}, even though oxygen is a little
heavier than nitrogen, the neon configuration oxygen approaches after
gaining 2 electrons intimate more stability than that of
N\textsuperscript{1+}. The ultimate aim of the addition of orbitals
(i.e., basis functions) is to allow an optimal reorganization of the
electronic cloud in the solid environment as compared to atomic or
ionic ones. 

Computational details germane to a replication of our work follow.
Ferroelectric sodium nitrite (ferro-NaNO\-\textsubscript{2}) has an
orthorhombic structure [29]. It is in the space group 
$\mathrm{C_{2v}^{20}}$ -- IMM2, with space group number 44 and Patterson symmetry \textit{Immm} [30].
The ferro-NaNO\textsubscript{2} unit cell contains eight atoms: four
(4) cations and four (4) anions whose positions are as indicated
between parentheses: Na: (0,0.5881,0), (0.5,0.0881,0.5); N: (0,0.1224,0), (0.5,0.6224,0.5); and O: (0,0,0.1962), 
(0.5,0.5,0.6962), (0,0,0.8038), (0.5,0.5,0.3038). 

We carried out five (5) different calculations utilizing five (5)
different basis sets in search of the optimal one. Table 2 contains the
basis sets utilized in the five (5) self-consistent calculations. 
Methodical increases of the basis set led to calculation IV as the one
with the optimal basis set; i.e., the calculation yielding the minima
of all the occupied energies. Calculation V does not lower any occupied
energies as compared to Calculation IV. Hence, the electronic structure
and related properties presented here were obtained with basis set IV,
the optimal basis set.

Our self-consistent computations were performed utilizing experimental
lattice constants of 3.518 {\AA}, 5.535 {\AA} and 5.382 {\AA} for
\textit{a}, \textit{b}, and \textit{c}, respectively, as obtained at 120
\textit{K} [31]. Neutral charge calculations (Na\textsuperscript{0},
N\textsuperscript{0} and O\textsuperscript{0}) were carried out. Calculated
charges were found to be +1 for Na, +1 for N and -2 for O. Then, we
performed self-consistent calculations for Na\textsuperscript{1+},
N\textsuperscript{1+} and O\textsuperscript{2-} to get the input
quantities for the calculations for NaNO\textsubscript{2}. The radial
parts of the atomic wave functions were expanded in terms of Gaussian
functions, employing a set of even-tempered Gaussian exponents.

The self-consistent atomic calculations provided trial atomic potentials
for Na, N, and O, respectively. These potentials were used to construct
the initial potential for ferro-NaNO\textsubscript{2}. We used 16
Gaussian functions for the \textit{s} and \textit{p} states and 14 for
the \textit{d} states for Na and N, respectively. We utilized 17
Gaussian functions for the \textit{ s} and \textit{p} states and 15 for
the \textit{d} states for O. A mesh of 48 \textit{k} points, with
proper weights in the irreducible Brillouin zone, was employed in the
self-consistent calculations. A total of 121
weighted k-points were used in band structure 
calculations and a total of 147 weighted k-points were employed to 
generate the energy eigenvalues for the electronic density of state
calculation. The computational error for the valence charge was about
0.073343 for 52 electrons, a little more than 10\textsuperscript{{}-3}
per electron. The self-consistent potentials converged to a difference
around 10\textsuperscript{{}-5} after about 60 iterations.

\section{Calculated Electronic Structure}
Our calculated, ab-initio, self-consistent bands for
ferro-NaNO\textsubscript{2} (Calculation IV) are as shown in Fig. 1 and
the comparison plot of the electron energy bands for basis set IV
(solid lines) and basis set V (dashed lines) are as shown in Fig. 2. As
is apparent in this graph, the eigenvalues of the occupied states
totally converged, within computational errors of about 0.050 eV,
showing clearly that Calculation IV is the one with the optimal basis
set. 
\bigskip
\bigskip

\includegraphics[width=3.2083in,height=2.278in]{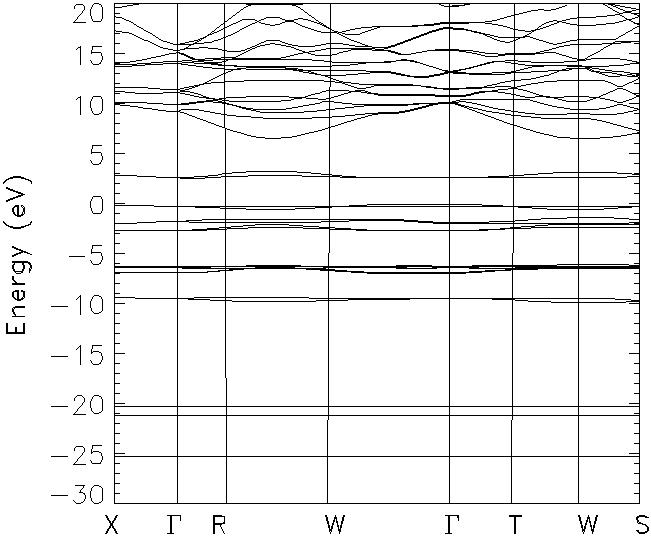}

{\centering
\noindent \textbf{Fig. 1.} The calculated electronic energy bands of
ferro-NaNO\textsubscript{2}, from Calculation IV.}

\bigskip

\bigskip

\includegraphics[width=3.1563in,height=2.4071in]{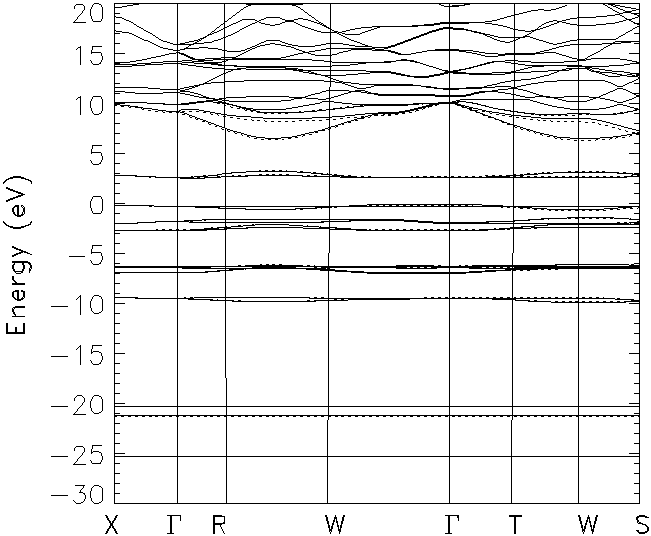}

\noindent \textbf{Fig. 2.} The calculated electronic energy bands of
ferro-NaNO\textsubscript{2}, from Calculations IV (full lines) and V
(dashed lines). The occupied energies from Calculations IV and V are
practically identical.

\includegraphics[width=3.75in,height=2.9744in]{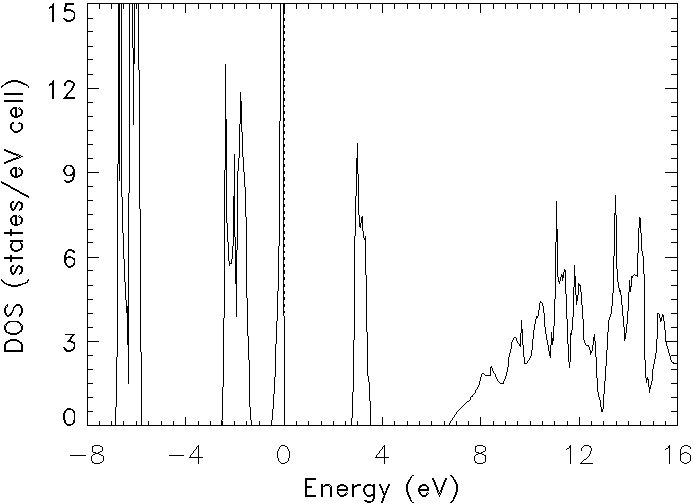}

\noindent \textbf{Fig. 3.} The calculated density of states (DOS) of
ferro-NaNO\textsubscript{2}, obtained with the BZW optimal basis set
(i.e. Calculation IV).

\bigskip
\bigskip

\includegraphics[width=4.028in,height=3.1807in]{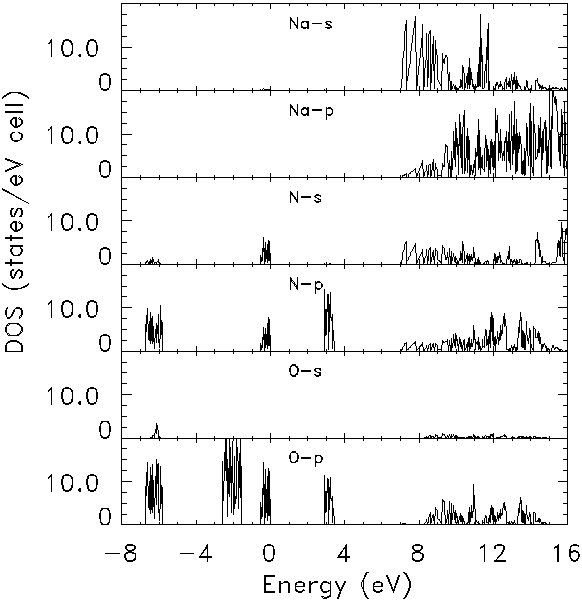}

\noindent \textbf{Fig. 4.} The calculated partial density of states (pDOS) of
ferro-NaNO\textsubscript{2} as obtained with the BZW optimal basis set
(i.e. Calculation IV).

\clearpage
\noindent \textbf{Table 1}. Search for the optimal basis sets (Orbital added is
in bold), as per the BZW method, for the description of the valence
states of ferroelectric sodium nitrite (ferro-NaNO\textsubscript{2}).
The optimal basis set is that from Calculation IV. Na is Sodium, N is
Nitrogen, and O is Oxygen. The maximum of the Valence band occurred at
\textit{W, }while the minimum of the conduction band occurred at
\textit{R.} NaNO\textsubscript{2 }is an indirect band gap material.

\begin{center}
\tablehead{}\begin{supertabular}{|m{0.44205984in}|m{0.6087598in}|m{0.8170598in}|m{0.6504598in}|m{0.6087598in}|m{0.7962598in}|m{0.7545598in}|m{0.67815983in}|}
\hline
\centering \bfseries Basis Set &
\centering  \textbf{Na, N and
O}\textbf{\textsubscript{2}}\textbf{ Core} &
\centering \bfseries Na Valence &
\centering \bfseries N \ Valence &
\centering  \textbf{O}\textbf{\textsubscript{2
}}\textbf{Valence} &
\centering \bfseries Total No. of Valence
orbitals &
\multicolumn{2}{m{1.5114598in}|}{\centering
\bfseries Band Gap (eV)}\\\hline
 &
 &
 &
 &
 &
 &
\centering \bfseries Direct at ${\Gamma}$ &
\centering\arraybslash \bfseries
Indirect\\\hline
\centering I &
 1s &
 2s2p3s &
 2s2p &
 2s2p &
\centering  34 &
\centering  2.40 &
\centering\arraybslash  2.38\\\hline
\centering  II &
 1s &
 2s2p3s\textbf{3p} &
 2s2p &
 2s2p &
\centering  40 &
\centering  2.70 &
\centering\arraybslash 2.53\\\hline
\centering  III &
 1s &
 2s2p3s3p &
 2s2p\textbf{3s} &
 2s2p &
\centering  42 &
\centering  3.18 &
\centering\arraybslash  2.93\\\hline
\centering \itshape\color{blue} IV &
\itshape\color{blue} 1s &

\textit{\textcolor{blue}{2s2p3s3p}}\textbf{\textit{\textcolor{blue}{3d}}}
&
\itshape\color{blue} 2s2p3s &
\itshape\color{blue} 2s2p &
\centering \itshape\color{blue} 52 &
\centering \itshape\color{blue} 3.22 &
\centering\arraybslash \itshape\color{blue}
2.83\\\hline
\centering\color{black}   V &
 1s &
 2s2p3s3p3d &
\color{black}  2s2p3s &
2s2p\textbf{3s} &
\centering  56 &
\centering  3.24 &
\centering\arraybslash  2.83\\\hline
\end{supertabular}
\end{center}

\bigskip
\bigskip

\noindent \color{black} \textbf{Table 2.} The Calculated Eigenvalues (in eV) at the High
Symmetry Points for ferro-NaNO\textsubscript{2}. The eigenvalues are
obtained by setting the energy at the top of the valence band, which
occurred at \textit{W, }equal to zero.

\color{black} \begin{center}
\tablehead{}\begin{supertabular}{|m{0.6698598in}|m{0.6441598in}|m{0.6441598in}|m{0.6441598in}|m{0.6698598in}|m{0.6441598in}|m{0.6698598in}|m{0.6511598in}|}
\hline
\multicolumn{8}{|m{5.7885594in}|}{\centering 
\color{black} X}\\\hline
\raggedleft \color{black} {}-25.11 &
\raggedleft \color{black} {}-25.11 &
\raggedleft \color{black} {}-21.03 &
\raggedleft \color{black} {}-21.03 &
\raggedleft \color{black} {}-21.01 &
\raggedleft \color{black} {}-21.01 &
\raggedleft \color{black} {}-20.97 &
\raggedleft\arraybslash \color{black}
{}-20.97\\\hline
\raggedleft \color{black} {}-20.05 &
\raggedleft \color{black} {}-20.05 &
\raggedleft \color{black} {}-9.24 &
\raggedleft \color{black} {}-9.24 &
\raggedleft \color{black} {}-6.73 &
\raggedleft \color{black} {}-6.73 &
\raggedleft \color{black} {}-6.21 &
\raggedleft\arraybslash \color{black}
{}-6.21\\\hline
\raggedleft \color{black} {}-6.08 &
\raggedleft \color{black} {}-6.08 &
\raggedleft \color{black} {}-2.51 &
\raggedleft \color{black} {}-2.51 &
\raggedleft  \textcolor{black}{{}-1.80} &
\raggedleft  \textcolor{black}{{}-1.80} &
\raggedleft \color{black} {}-0.01 &
\raggedleft\arraybslash \color{black}
{}-0.01\\\hline
\raggedleft \color{black} 3.01 &
\raggedleft \color{black} 3.01 &
\raggedleft \color{black} 10.08 &
\raggedleft \color{black} 10.08 &
\raggedleft \color{black} 10.36 &
\raggedleft \color{black} 10.36 &
\raggedleft \color{black} 11.38 &
\raggedleft\arraybslash \color{black}
11.38\\\hline
\raggedleft \color{black} 11.79 &
\raggedleft \color{black} 11.79 &
\raggedleft \color{black} 13.96 &
\raggedleft \color{black} 13.96 &
\raggedleft \color{black} 14.12 &
\raggedleft \color{black} 14.12 &
\raggedleft \color{black} 14.32 &
\raggedleft\arraybslash \color{black}
14.32\\\hline
\raggedleft \color{black} 17.49 &
\raggedleft \color{black} 17.49 &
\raggedleft  \textcolor{black}{18.90} &
\raggedleft  \textcolor{black}{18.90} &
\raggedleft \color{black} 19.82 &
\raggedleft \color{black} 19.82 &
\raggedleft \color{black} 24.92 &
\raggedleft\arraybslash \color{black}
24.92\\\hline
\raggedleft \color{black} 30.26 &
\raggedleft \color{black} 30.26 &
~
 &
~
 &
~
 &
~
 &
~
 &
~
\\\hline
\multicolumn{8}{|m{5.7885594in}|}{\centering 
${\Gamma}$}\\\hline
\raggedleft \color{black} {}-25.11 &
\raggedleft \color{black} {}-25.09 &
\raggedleft \color{black} {}-21.13 &
\raggedleft \color{black} {}-21.03 &
\raggedleft \color{black} {}-21.02 &
\raggedleft \color{black} {}-21.01 &
\raggedleft \color{black} {}-20.89 &
\raggedleft\arraybslash \color{black}
{}-20.89\\\hline
\raggedleft \color{black} {}-20.13 &
\raggedleft \color{black} {}-20.07 &
\raggedleft \color{black} {}-9.68 &
\raggedleft \color{black} {}-9.28 &
\raggedleft \color{black} {}-6.34 &
\raggedleft \color{black} {}-6.32 &
\raggedleft \color{black} {}-6.3 &
\raggedleft\arraybslash \color{black}
{}-6.16\\\hline
\raggedleft 
\textcolor{black}{{}-6}\textcolor{black}{.00} &
\raggedleft \color{black} {}-5.89 &
\raggedleft \color{black} {}-2.04 &
\raggedleft \color{black} {}-1.76 &
\raggedleft  \textcolor{black}{{}-1.70} &
\raggedleft \color{black} {}-1.33 &
\raggedleft \color{black} {}-0.51 &
\raggedleft\arraybslash 
\textcolor{black}{{}-0.10}\\\hline
\raggedleft \color{black} 3.12 &
\raggedleft \color{black} 3.51 &
\raggedleft \color{black} 6.66 &
\raggedleft \color{black} 8.59 &
\raggedleft \color{black} 9.31 &
\raggedleft \color{black} 9.61 &
\raggedleft \color{black} 10.31 &
\raggedleft\arraybslash \color{black}
10.91\\\hline
\raggedleft \color{black} 12.53 &
\raggedleft \color{black} 13.59 &
\raggedleft \color{black} 13.86 &
\raggedleft \color{black} 13.97 &
\raggedleft \color{black} 13.97 &
\raggedleft \color{black} 14.67 &
\raggedleft \color{black} 14.78 &
\raggedleft\arraybslash \color{black}
16.26\\\hline
\raggedleft \color{black} 16.47 &
\raggedleft \color{black} 18 &
\raggedleft \color{black} 18.84 &
\raggedleft \color{black} 20.17 &
\raggedleft \color{black} 20.43 &
\raggedleft \color{black} 20.86 &
\raggedleft \color{black} 20.96 &
\raggedleft\arraybslash \color{black}
21.97\\\hline
\raggedleft \color{black} 22.06 &
\raggedleft \color{black} 31.43 &
~
 &
~
 &
~
 &
~
 &
~
 &
~
\\\hline
\multicolumn{8}{|m{5.7885594in}|}{\centering 
R}\\\hline
\raggedleft \color{black} {}-25.15 &
\raggedleft \color{black} {}-25.15 &
\raggedleft  \textcolor{black}{{}-21.00} &
\raggedleft  \textcolor{black}{{}-21.00} &
\raggedleft \color{black} {}-20.99 &
\raggedleft \color{black} {}-20.99 &
\raggedleft \color{black} {}-20.92 &
\raggedleft\arraybslash \color{black}
{}-20.92\\\hline
\raggedleft \color{black} {}-20.18 &
\raggedleft \color{black} {}-20.18 &
\raggedleft \color{black} {}-9.44 &
\raggedleft \color{black} {}-9.44 &
\raggedleft \color{black} {}-6.74 &
\raggedleft \color{black} {}-6.74 &
\raggedleft \color{black} {}-6.14 &
\raggedleft\arraybslash \color{black}
{}-6.14\\\hline
\raggedleft \color{black} {}-6.08 &
\raggedleft \color{black} {}-6.08 &
\raggedleft \color{black} {}-2.38 &
\raggedleft \color{black} {}-2.38 &
\raggedleft \color{black} {}-1.47 &
\raggedleft \color{black} {}-1.47 &
\raggedleft \color{black} {}-0.07 &
\raggedleft\arraybslash \color{black}
{}-0.07\\\hline
\raggedleft \color{black} 2.83 &
\raggedleft \color{black} 2.83 &
\raggedleft \color{black} 9.18 &
\raggedleft \color{black} 9.18 &
\raggedleft \color{black} 10.12 &
\raggedleft \color{black} 10.12 &
\raggedleft \color{black} 11.02 &
\raggedleft\arraybslash \color{black}
11.02\\\hline
\raggedleft \color{black} 12.02 &
\raggedleft \color{black} 12.02 &
\raggedleft \color{black} 13.4 &
\raggedleft \color{black} 13.4 &
\raggedleft \color{black} 14.46 &
\raggedleft \color{black} 14.46 &
\raggedleft \color{black} 15.62 &
\raggedleft\arraybslash \color{black}
15.62\\\hline
\raggedleft \color{black} 15.73 &
\raggedleft \color{black} 15.73 &
\raggedleft \color{black} 17.96 &
\raggedleft \color{black} 17.96 &
\raggedleft \color{black} 22.35 &
\raggedleft \color{black} 22.35 &
\raggedleft \color{black} 24.19 &
\raggedleft\arraybslash \color{black}
24.19\\\hline
\raggedleft \color{black} 24.99 &
\raggedleft \color{black} 24.99 &
~
 &
~
 &
~
 &
~
 &
~
 &
~
\\\hline
\multicolumn{8}{|m{5.7885594in}|}{\centering 
W}\\\hline
\raggedleft \color{black} {}-25.15 &
\raggedleft \color{black} {}-25.15 &
\raggedleft \color{black} {}-21.02 &
\raggedleft \color{black} {}-21.02 &
\raggedleft  \textcolor{black}{{}-21.00} &
\raggedleft  \textcolor{black}{{}-21.00} &
\raggedleft \color{black} {}-20.98 &
\raggedleft\arraybslash \color{black}
{}-20.98\\\hline
\raggedleft \color{black} {}-20.06 &
\raggedleft \color{black} {}-20.06 &
\raggedleft \color{black} {}-9.28 &
\raggedleft \color{black} {}-9.28 &
\raggedleft \color{black} {}-6.84 &
\raggedleft \color{black} {}-6.84 &
\raggedleft \color{black} {}-6.13 &
\raggedleft\arraybslash \color{black}
{}-6.13\\\hline
\raggedleft \color{black} {}-6.08 &
\raggedleft \color{black} {}-6.08 &
\raggedleft \color{black} {}-2.41 &
\raggedleft \color{black} {}-2.41 &
\raggedleft \color{black} {}-1.79 &
\raggedleft \color{black} {}-1.79 &
\raggedleft \color{black} 0 &
\raggedleft\arraybslash \color{black} 0\\\hline
\raggedleft \color{black} 2.98 &
\raggedleft \color{black} 2.98 &
\raggedleft \color{black} 10.31 &
\raggedleft \color{black} 10.31 &
\raggedleft \color{black} 10.39 &
\raggedleft \color{black} 10.39 &
\raggedleft \color{black} 10.93 &
\raggedleft\arraybslash \color{black}
10.93\\\hline
\raggedleft \color{black} 11.59 &
\raggedleft \color{black} 11.59 &
\raggedleft \color{black} 13.38 &
\raggedleft \color{black} 13.38 &
\raggedleft  \textcolor{black}{13.50} &
\raggedleft  \textcolor{black}{13.50} &
\raggedleft \color{black} 15.72 &
\raggedleft\arraybslash \color{black}
15.72\\\hline
\raggedleft \color{black} 17.68 &
\raggedleft \color{black} 17.68 &
\raggedleft \color{black} 18.34 &
\raggedleft \color{black} 18.34 &
\raggedleft \color{black} 19.96 &
\raggedleft \color{black} 19.96 &
\raggedleft \color{black} 24.09 &
\raggedleft\arraybslash \color{black}
24.09\\\hline
\raggedleft \color{black} 26.91 &
\raggedleft \color{black} 26.91 &
~
 &
~
 &
~
 &
~
 &
~
 &
~
\\\hline
\multicolumn{8}{|m{5.7885594in}|}{\centering 
T}\\\hline
\raggedleft \color{black} {}-25.14 &
\raggedleft \color{black} {}-25.14 &
\raggedleft \color{black} {}-21.08 &
\raggedleft \color{black} {}-21.08 &
\raggedleft \color{black} {}-21.01 &
\raggedleft \color{black} {}-21.01 &
\raggedleft \color{black} {}-20.89 &
\raggedleft\arraybslash \color{black}
{}-20.89\\\hline
\raggedleft \color{black} {}-20.11 &
\raggedleft \color{black} {}-20.11 &
\raggedleft \color{black} {}-9.53 &
\raggedleft \color{black} {}-9.53 &
\raggedleft \color{black} {}-6.31 &
\raggedleft \color{black} {}-6.31 &
\raggedleft \color{black} {}-6.15 &
\raggedleft\arraybslash \color{black}
{}-6.15\\\hline
\raggedleft \color{black} {}-6.06 &
\raggedleft \color{black} {}-6.06 &
\raggedleft \color{black} {}-1.91 &
\raggedleft \color{black} {}-1.91 &
\raggedleft \color{black} {}-1.52 &
\raggedleft \color{black} {}-1.52 &
\raggedleft \color{black} {}-0.23 &
\raggedleft\arraybslash \color{black}
{}-0.23\\\hline
\raggedleft \color{black} 3.28 &
\raggedleft \color{black} 3.28 &
\raggedleft \color{black} 7.33 &
\raggedleft \color{black} 7.33 &
\raggedleft \color{black} 9.74 &
\raggedleft \color{black} 9.74 &
\raggedleft \color{black} 10.95 &
\raggedleft\arraybslash \color{black}
10.95\\\hline
\raggedleft \color{black} 12.71 &
\raggedleft \color{black} 12.71 &
\raggedleft \color{black} 13.11 &
\raggedleft \color{black} 13.11 &
\raggedleft \color{black} 14.43 &
\raggedleft \color{black} 14.43 &
\raggedleft \color{black} 16.59 &
\raggedleft\arraybslash \color{black}
16.59\\\hline
\raggedleft \color{black} 17.99 &
\raggedleft \color{black} 17.99 &
\raggedleft \color{black} 18.98 &
\raggedleft \color{black} 18.98 &
\raggedleft  \textcolor{black}{19.60} &
\raggedleft  \textcolor{black}{19.60} &
\raggedleft \color{black} 20.71 &
\raggedleft\arraybslash \color{black}
20.71\\\hline
\raggedleft 
\textcolor{black}{25}\textcolor{black}{.00} &
\raggedleft  \textcolor{black}{25.00} &
~
 &
~
 &
~
 &
~
 &
~
 &
~
\\\hline
\end{supertabular}
\end{center}

\bigskip

Table 2 shows the energies at the high symmetry points. Figs. 3 and 4
show the calculated, total (DOS) and partial (pDOS) density of states
derived from the bands in Fig. 1. The calculated peaks in the valence
band DOS closest to the Fermi energy are at -0.25 $\pm$ 0.1 eV and 
-2.4 $\pm$ 0.2 eV. We found a third peak at -5.8 $\pm$ 0.1 eV which agrees rather 
well with the experimental value of -5.8 eV
reported by Kamada et al. [32]. In the conduction band, we found
relatively broad peaks whose centers are located at 8.25, 9.50, 10.50,
11.25, and 12.00 eV, respectively. The widths of these broad peaks are
about 0.5 eV. Caution is required in using the conduction band density
of states at energies above 6 eV. As explained by Jin et al. [33] and
Bagayoko et al. [24, 34], BZW-EF calculations are concerned with correctly
solving equations describing the ground states. Hence, while the
low-laying excited states from these calculations have been found to
agree with measurement, upper excited states are not expected to be
correctly described. Specifically, the calculated optical properties of
wurtzite InN [33] agree with experiment up to energies of 5.5 to 6.0
eV. The comparison of the bands from Calculations IV and V in Fig. 2
suggests that our calculated, excited states for NaNO\textsubscript{2}
could be meaningful up to 20 eV. This situation could be a feature of
molecular solids.

Our calculated electron effective masses in the ${\Gamma}$-X,
${\Gamma}$--R, and ${\Gamma}$-W directions are 1.22, 0.64, and 0.74
m\textsubscript{o}, respectively. These values clearly corroborate the
anisotropic nature of this material.

Our calculations were performed at lattice constants for which
experimental data are available for comparison purposes. However, the
question arises as to what results could be obtained with a fully
optimized crystal structure (i.e., equilibrium lattice structure).
\ The optimization led to equilibrium lattice parameters of 3.529
{\AA}, 5.560 {\AA} and 5.391 {\AA} for \textit{a}, \textit{b} and
\textit{c}, respectively. The difference between our former (120 K
experimental lattice parameter) and the latter (with optimized lattice
parameters) eigenvalues is approximately 0.02 eV, with the energies
obtained with the experimental lattice constants being lower. The two
results are practically the same as their difference is smaller than
our computational uncertainty of 50 meV (0.05 eV).

\section{Discussion}

An added dimension of the validity of the work reported here stems from
the rigor of the BZW-EF method. Our calculations solved self-consistently
the system of equations defining LDA. They are totally ab-initio and do
not entail the use of two different theories; the sophisticated work of
Zakharov et al. [35] utilized LDA wave functions as input in their
quasi-particle calculations.

Unlike single trial basis set (STBS) DFT and other calculations, LDA
BZW-EF results agree with experiment and point to much more physical
content of LDA eigenvalues than previously thought. As explained by
Zhao et al. [23], STBS calculations do not avoid a well-defined
{\textquotedblleft}\textit{basis set and variational
effect} that has plagued electronic structure
calculations since their inception, with emphasis on DFT calculations.

Indeed, BZW-EF calculations begin with the minimum
basis set that is just large enough to account for all the electrons in
the system, i.e., NaNO\textsubscript{2} for this
work. It then augments the basis set methodically [22-24] and carries
out successive, self-consistent calculations with increasingly larger
basis sets. Except for the first calculation, the
occupied eigenvalues of
a calculation are compared numerically and graphically to those of the
calculation immediately preceding it. Ultimately, the
occupied eigenvalues of
two consecutive calculations, say N
and (N+1), are found to be identical-within computational
uncertainties of about 50 meV. The results from calculation
N provide the physical
description of the system and the corresponding basis set is the
optimal basis set. While Calculation (N+1) yields the same occupied energies as
Calculation N, some of
the unoccupied energies from it are generally lower than corresponding
ones from Calculation N.
This extra-lowering of the unoccupied eigenvalues,
for calculations with basis sets larger than the optimal one, is due to
a basis set and
variational effect. It is a
mathematical artifact stemming from the Rayleigh theorem. For basis
sets larger than the optimal one, the resulting charge density and
potential are identical to those from Calculation N [23]. So, the above
extra-lowering is not due to a physical interaction.

From the density of states plots, our calculated peaks of -0.25 $\pm$ 0.1 eV
 and -2.4 $\pm$ 0.2 eV in the valence band DOS agree with the x-ray photoelectron results of
Calabrese and Heyes, [36] and the theoretical calculations of Ravindran
et al. [3] and Zhuravlev and Korabel{\textquoteright}nikov [12]. The
observed sharp peak at -0.25 eV arises mainly from the hybridization
between the O \textit{2p }, N \textit{2p}, and N \textit{2s}. The peak 
at -2.4 $\pm$ 0.2 eV arises from the O \textit{2p} orbitals. These observations are in
agreement with that of Zhuravlev and Korabel{\textquoteright}nikov [12]
that the upper valence bands are formed by complex anion states. In the
conduction bands, our calculated peaks of 3.25
eV, 8.25 eV, 9.5 eV, 10.50 eV and 11.25 eV compare
favorably with the room
temperature vacuum ultraviolet (VUV) data
from measured values
of Yamashita and Kato [37] and Ashida et al. [38],
respectively.

A distinct feature of our calculated band structure in Figure 1 is the
flat nature of the top of the valence band as well as that of the
bottom of the conduction band. This singular feature means that the
band structure of ferro-NaNO\textsubscript{2} can well be described by
localized molecular levels of an ionic crystal. This observation is
similar to that of Ravindran et al. [2], Wang et al. [11] and of Jiang
et al. [14]. Above the Fermi energy, as can be seen from the DOS, many
conduction bands exist and some are highly dispersed. From the pDOS for
the upper valence and the conduction bands, we can infer that the Na
states are high up in the conduction bands and do not partake in the
formation of the molecular solid.

The top-most valence bands and bottom-most conduction bands are flat.
Consequently, the exact maximum and minimum values of the valence and
conduction bands, respectively, cannot be easily identified by merely
looking at the plots. From our outputs, before setting the top of the
valence bands equal to zero, we found that the maximum of the valence
band (VBM) is at the \textit{W} point while the minimum of the conduction band (CBM) is at 
\textit{R}, respectively, giving a
fundamental, indirect energy band gap of 2.83 eV from Calculation IV.
Our calculated direct gaps are 2.90, 2.98, 3.02, 3.22, and 3.51 eV at
\textit{R, W, X, ${\Gamma}$}, and \textit{T}, respectively. The
scanning electron microscopy measurements of Balabinskaya et al. [9]
reported a direct band gap of 3.0 eV.

The extent to which our electronic
structure calculations take temperature into account is practically
limited to the use of lattice constants obtained at a given
temperature. With the lattice constants obtained at 120 K, our
calculated direct gap of 3.22 eV compares favorably
with  the infrared absorption result of 3.22 eV
obtained by Sidman [6] at a low temperature (77 K). Predictably, it is
larger than the room temperature (293.15 K) measurement of 3.14 eV of
Verkhovskaya and Sonin [7].

The positions of our VBM (at \textit{W}) and CBM (at \textit{R}) are the
same as those from the work of Ravindran et al. [11] that found a
calculated, indirect band gap of 2.2 eV. Jiang et al. [14] calculated
an indirect gap of 2.95 eV. However, these authors obtained their VBM
at \textit{S}. In our calculations, only the \textit{1s }states of
\textit{Na, N}, and \textit{O}, respectively, were in the core. Jiang
et al. [14] placed \textit{1s, 2s}, and \textit{2p} electrons of sodium
in the core.

 Our calculated electron effective masses of 1.18, 0.63, and 0.73
m\textsubscript{o }in the ${\Gamma}$-X, ${\Gamma}$--R, and
${\Gamma}$-W directions, respectively, show the highly anisotropic
nature of this material. We know of no experimental or calculated
electron effective masses for NaNO\textsubscript{2}, except the
semi-empirical, temperature dependent result of El-Dib and Hassan
[8]\textsuperscript{ }who found an electron effective mass of 0.61
m\textsubscript{o} in the ${\Gamma}$--R direction. Our result of 0.63,
from ${\Gamma}$--R, agrees with their finding.

\section{Conclusion}

The above results and related discussions are expected to add to our
understanding of ferro-NaNO\textsubscript{2}. The noted agreement
between our calculated results and experiment will hopefully motivate
further experimental and theoretical studies of this material. Our
calculated eigenvalues at high symmetry points should aid in some
comparison with future studies.

This work and similar ones from this group [17-24] indicate that LDA can
correctly describe and predict electronic and related properties of
semiconductors, provided one methodically search for the optimal basis
set that minimizes the occupied energies. The predictive capability of
such calculations is expected to inform and to guide the design and
fabrication of semiconductor based devices. For example, for binary
to quaternary systems, theoretical studies of variations of the
concentration of an element (or more) have the potential to impact
positively industrial activities.

\bigskip
\section*{Acknowledgments}

This work was funded in part by the National Science Foundation and the
Louisiana Board of Regents (Award Nos. EPS-1003897, and NSF
(2010-15)-RII-SUBR, 0754821, and HRD-1002541), the Louisiana Optical
Network Initiative (LONI, SUBR Award No. 2-10915). CEE wishes to thank
Govt. of Ebonyi State, Nigeria.\newline

\section*{References}

\liststyleWWviiiNumii
\begin{enumerate}
\item {
S. Sawada, S. Nomura, S. Fuji, and I. Yoshida, Phys. Rev. Letts.
\textbf{1}, 320 (1958).}
\item {
P. Ravindran, A. Delin, B. Johansson, and O. Eriksson, Phys. Rev B
\textbf{59}, 1776 (1999).}
\item {
M. Kamada, M. Yoshikawa, and R. Kato, J. Phys. Soc. Jpn. \textbf{39},
1004 (1975).}
\item {
Y. Asao, I. Yoshida, R. Ando, and S. Sawada, J. Phys. Soc. Jpn.,
\textbf{17}, 442 (1962). \ }
\item {
Y. Takagi and K. Gesi, J. Phys. Soc. Jpn, \textbf{22}, 979 (1967).}
\item {
J.W. Sidman, J. Am. Chem. Soc. \textbf{79}, 2669 (1957).}
\item {
\foreignlanguage{swedish}{K.A. Verkhovskaya and A.S. Sonin, Soc.
}Phys-JETP \textbf{25}, 249 (1967).}
\item {
A.M. El-Dib and H.F. Hassan, Phys. Stat. Sol. B \textbf{126}, 587
(1984).}
\item {
A. S. Balabinskaya, E. N. Ivanova, M. S. Ivanova, Yu. A. Kumzerov, S. V.
Pan{\textquoteright}kova, V. V. Poborchii, S. G. Romanov, V. G.
Solovyev, and S. D. Khanin, Glass Phys. and Chem. \textbf{31} (3),
330-336 (2005).}
\item {
K. --S Kam and J.H. Henkel, Phys. Rev. B \textbf{17}, 1361 (1978).}
\item {
Y.X. Wang, W.L. Zhong, C.L. Wang, P.L. Zhang, and Y.P. Peng, Phys.
Letts. A \textbf{269}, 252--256 (2000); ibid, Solid State Comm.
\textbf{112} 495--498 (1999). }
\item {
Y. N. Zhuravlev and D. V. Korabel{\textquoteright}nikov, Phys. of the
Solid State, \textbf{51 }(1), 69-77 (2009).}
\item {
Y.N. Zhuravlev and A. S. Poplavnoi, Russian Phys. J., \textbf{44} (12)
(2001).}
\item {
H. Jiang,~Y.N. Xu, and W. Y. Ching, Ferroelectrics, \textbf{136} (1),
137-146 (1992).}
\item {
Maureen I. McCarthy, K. A. Peterson, \ and W.P. Hess, J. Phys. Chem.
\textbf{100}\textit{, }6708-6714 (1996).}
\item {
J. H. Henkel\textbf{\textit{\textsuperscript{ }}}, T. C. Collins, J. L.
Iveyt, and R. N. Euwema, Int. J. of Quant. Chem. \textbf{9},
\textstyleStrongEmphasis{\textmd{529~--~533 (2009).}}}
\item {
D Bagayoko , G L Zhao , J D Fan, and J T Wang, \textstyleEmphasis{J.
Phys.: Condens. Matter} \textbf{10} 5645 (1998).}
\item {
E. C. Ekuma and D. Bagayoko, Jpn. J. Appl. Phys. \textbf{50}, 101103
(2011). \ \ }
\item {
E. C. Ekuma, L. Franklin, J. T. Wang, G. L. Zhao, and D. Bagayoko,
\textit{Can. J. Phys. }\textbf{\textit{89}}\textit{ (3), pg. 319-324
}(2011). See also: E. C. Ekuma, L. Franklin, J. T. Wang, G. L. Zhao,
and D. Bagayoko, \textit{Physica B, }\textbf{\textit{406}}\textit{ (8),
pg. 1477-1480 }(2011).}
\item {
D. Bagayoko and G. L. Zhao, Physica C 364-365, Pages 261-264 (2001).}
\item {
D. Bagayoko, L. Franklin, and G. L. Zhao, J. Appl.
\foreignlanguage{ngerman}{Phys. 96, 4297-4301 (2004); }D. Bagayoko,
\ L. Franklin, G. L. Zhao, and H. Jin, \ J. Appl. Phys. 103, 096101
(2008).}
\item {
C. E. Ekuma, {M. Jarrell, J. Moreno, and D.
Bagayoko, AIP Advances
}{\textbf{2}}{,
}012189 (2012).}
\item {
G. L. Zhao, D. Bagayoko, and T. D. Williams. Physical Review B60, 1563,
1999.}
\item {
D. Bagayoko. Proceedings, International Seminar on Theoretical Physics
and Applications to Development (ISOTPAND), August 2008, Abuja,
Nigeria. \textcolor{black}{Available in the African Journal of Physics
(}\url{http://sirius-c.ncat.edu/asn/ajp/allissue/ajp-ISOTPAND/index.html}\textcolor{black}{).}}
\item {
D. M. Ceperley and B. J. Alder, Phys. Rev. Lett. 45, 566 (1980).}
\item {
S. H. Vosko, L. Wilk, and M. Nusair, Can. J. Phys. 58, 1200 (1980).}
\item {
B. N. Harmon, W. Weber, and D. R. Hamann, Phys. Rev. B 25, 1109 (1982).}
\item {
W. Kohn and L. J. Sham, Phys. Rev. 140, A1133 (1965).}
\item {
\textcolor{black}{G.E. Zeigler, Phys. Rev.
}\textbf{\textcolor{black}{38}}\textcolor{black}{, 1040 (1931).}}
\item {
International Tables for Crystallography, Vol. A: Space-Group Symmetry,
5\textsuperscript{th} Edn. Theo Hahn, Ed. Springer (2005).}
\item {
M. Okuda, S. Ohba, Y. Saito, T. Ito, and I shibya, Acta Cryst. B
\textbf{46}, 343 (1990) in, as quoted in Inorganic Crystal Structure
Database (ICSD), NIST, Release 2010/1.}
\item {
\textcolor{black}{M. Kamada, K. Ichikawa and K. Tsutsumi, J. Phys. Soc.
Jpn. }\textbf{\textcolor{black}{50 }}\textcolor{black}{(1), 170
(1981).}}
\item {
H. Jin, G. L. Zhao, and D. Bagayoko, J. Appl. Phys. 101, 033123 (2007).}
\item {
D. Bagayoko, L. Franklin, and G. L. Zhao, Phys. Rev. B \textbf{76},
037101 (2007)}
\item {
O. Zakharov, A. Rubio, X. Blase, M. Cohen, and S. Louie, Phys. Rev. B
\textbf{50}, 10780 (1994).}
\item {
A. Calabrese and R. G. Hayes\textcolor{black}{, J. Electron Spectrosc.
Relat. Phenom. }\textbf{\textcolor{black}{6}}\textcolor{black}{(1), 1
(1975).}}
\item {
A. Yamashita and R. Kato, \textcolor{black}{J. Phys. Soc. Jpn.
}\textbf{\textcolor{black}{29 }}\textcolor{black}{(6), 1557 (1970).}}
\item {
M. Ashida, O. Ohta, M. Kamada, M. Watanabe, and R. Kato,
\textcolor{black}{J. Electron Spectrosc. Relat. Phenom.
}\textbf{\textcolor{black}{79}}\textcolor{black}{, 55 (1996).}}
\end{enumerate}

\end{document}